\documentclass[a4paper]{spie}  

\usepackage{amsmath,amsfonts,amssymb}
\usepackage{graphicx}
\usepackage[colorlinks=true, allcolors=blue]{hyperref}
\usepackage{subcaption}
\usepackage{cleveref}
\newcommand*\diff{\mathop{}\!\mathrm{d}}
\newcommand*\Diff[1]{\mathop{}\!\mathrm{d^#1}}
\newcommand*\rme{\mathop{}\!\mathrm{e}}
\newcommand*\rmi{\mathop{}\!\mathrm{i}}
\newcommand*\crit{\mathop{}\!\mathrm{c}}
\newcommand*\Sl{\mathop{}\!\mathrm{S}}
\newcommand*\Sp{\mathop{}\!\mathrm{S'}}
\newcommand*\Spp{\mathop{}\!\mathrm{S''}}
\newcommand*\nn{\mathop{}\!\mathrm{n}}

\title{Proximity effect model for x-ray Transition Edge Sensors}

\author[a]{R. C. Harwin}
\author[a]{D. J. Goldie}
\author[a]{S. Withington}
\author[b]{P. Khosropanah}
\author[b]{L. Gottardi}
\author[b,c]{J.-R. Gao}
\affil[a]{Cavendish Laboratory, JJ Thomson Avenue, Cambridge, CB3 0HE, United Kingdom}
\affil[b]{SRON National Institute for Space Research, Sorbonnelaan 2, 3584 CA Utrecht, The Netherlands}
\affil[c]{Kavli Institute of NanoScience, Faculty of Applied Sciences, Delft University of Technology, Lorentzweg 1, 2628 CJ Delft, The Netherlands}

\authorinfo{Further author information: (Send correspondence to R.C.H.)\\R.C.H.: E-mail: rch66@cam.ac.uk}

\begin{document} 
\maketitle

\begin{abstract}
Transition Edge Sensors are ultra-sensitive superconducting detectors with applications in many areas of research, including astrophysics. The device consists of a superconducting thin film, often with additional normal metal features, held close to its transition temperature and connected to two superconducting leads of a higher transition temperature. There is currently no way to reliably assess the performance of a particular device geometry or material composition without making and testing the device. We have developed a proximity effect model based on the Usadel equations to predict the effects of device geometry and material composition on sensor performance. The model is successful in reproducing $I-V$ curves for two devices currently under study. We use the model to suggest the optimal size and geometry for TESs, considering how small the devices can be made before their performance is compromised. In the future, device modelling prior to manufacture will reduce the need for time-consuming and expensive testing.
\end{abstract}


\section{INTRODUCTION}
\label{sec:intro}
Transition Edge Sensors (TESs) are highly sensitive superconducting detectors, which are used or are being developed for use in a variety of astronomical instruments, including for the X-ray Integral Field Unit (X-IFU) on the ATHENA x-ray space telescope, \cite{Smith2016,Gottardi2016} the SAFARI spectrometer on the SPICA far infra-red telescope, \cite{Goldie2011,Goldie2016,Khosropanah2016,Williams2018} and the SPTPol experiment. \cite{Chang2012}

A TES uses the sharpness of the normal-resistive-superconducting transition to detect electromagnetic radiation \cite{Sadleir2010}, giving it excellent performance in terms of energy resolution when used as a microcalorimeter or noise equivalent power when operated as a bolometer. The device consists of a superconducting thin film (denoted as S') biased on its superconducting to normal-resistive transition, connected to superconducting electrical leads (denoted as S) of a material with a higher transition temperature. The transition temperature $T_{\crit,\Sp}$ of the thin film should be chosen to match experimental requirements determined by bath temperature and power handling, and so the film is typically fabricated as a bilayer such as Mo/Au, \cite{Goldie2012} Mo/Cu, \cite{Glowacka2010} or Ti/Au, \cite{Portesi2008} allowing precise control of $T_{\crit,\Sp}$ via the superconducting proximity effect.\cite{Martinis2000} Some TESs, such as those used to detect x-rays, may have an additional absorber on top of or next to the bilayer, typically made of a material with a large stopping power such as bismuth.

Arrays of TESs must be designed to meet the requirements of the relevant instrument. Desirable characteristics of a single device include low noise equivalent power, fast response time and small size, as smaller devices can be packed closely together in an array to produce good spatial resolution. We can use standard theoretical models to predict the effects of quantities such as heat capacity $C$ and device transition temperature $T_{\crit}$ on characteristics such as thermal response time and, for calorimeters, energy resolution. The response time of the device depends on the effective thermal time constant, $\tau_{\mathrm{eff}} \propto C/(G(1+\alpha/n))$, \cite{Irwin2005} where $G$ is the thermal conductance to the surrounding wafer. $\alpha = (T/R) (\partial R / \partial T)$ is a parameter that characterises the sharpness of the transition, with a sharper transition having a higher value of $\alpha$, and $n$ is a parameter that depends on the power flow to the surroundings. Therefore, to make the response time as fast as possible, a TES should have a low heat capacity, a high thermal conductance and a high value of $\alpha$. For a single TES when operated as a calorimeter, the energy resolution $\Delta E \propto T \sqrt{C k_{\mathrm{B}}/\alpha}$. \cite{Portesi2008} Here $T$ is the operating temperature of the device, typically close to the transition temperature $T_{\crit}$. To optimise energy resolution, calorimeters should therefore have low operating temperatures, small heat capacities and sharp transitions.

The thin film is often patterned with additional normal metal features (denoted $\Spp$) such as bars or dots to control edge effects, \cite{Glowacka2008} to modify the sharpness of the transition, \cite{Smith2008} and to reduce excess noise. \cite{Ullom2004} These normal metal features also reduce the transition temperature, \cite{Sadleir2011} increase the heat capacity of the bilayer and reduce the value of $\alpha$, \cite{Wang2015} and so it is often unclear as to their exact effect on the behaviour of the TES. We would like to identify the optimal number, position and size of normal metal features to optimise TES performance.

In general, regardless of the planned application, the device size also impacts design decisions, as the transition temperature and the heat capacity of the bilayer will vary with device size, which will in turn affect the device response time. The length of the device also affects the strength of the proximity effect from the superconducting electrodes to the bilayer. As well as this, for good spatial resolution, the TESs must be as tightly packed as possible to allow the focal plane to be sampled more densely, making smaller devices desirable. \cite{Smith2012} However, for some applications where large absorbers are necessary, the required absorber size will determine the maximum achievable spatial resolution. We are looking to find a device size that optimises both response time and spatial resolution, where possible.

The optimisation of TESs is typically carried out empirically rather than as part of the design process. We have developed a numerical model based on the Usadel equations and would like to be able to use this model to test proposed device designs prior to manufacture. By using this model to investigate the effects of normal metal bars and device size, our work could enable the selection of TES designs with the required performance parameters for a particular application. This would avoid the time and expense associated with the fabrication and testing of a large number of prototype TESs.

Here we describe our modelling process and compare results from the model with measured $I-V$ characteristics of representative devices. We have also used our model to investigate a series of square TESs with different side lengths, to study the variation of transition temperature with device size.

\section{METHOD}
\label{sec:method}

\begin{figure}
\centering
\includegraphics[width=\textwidth]{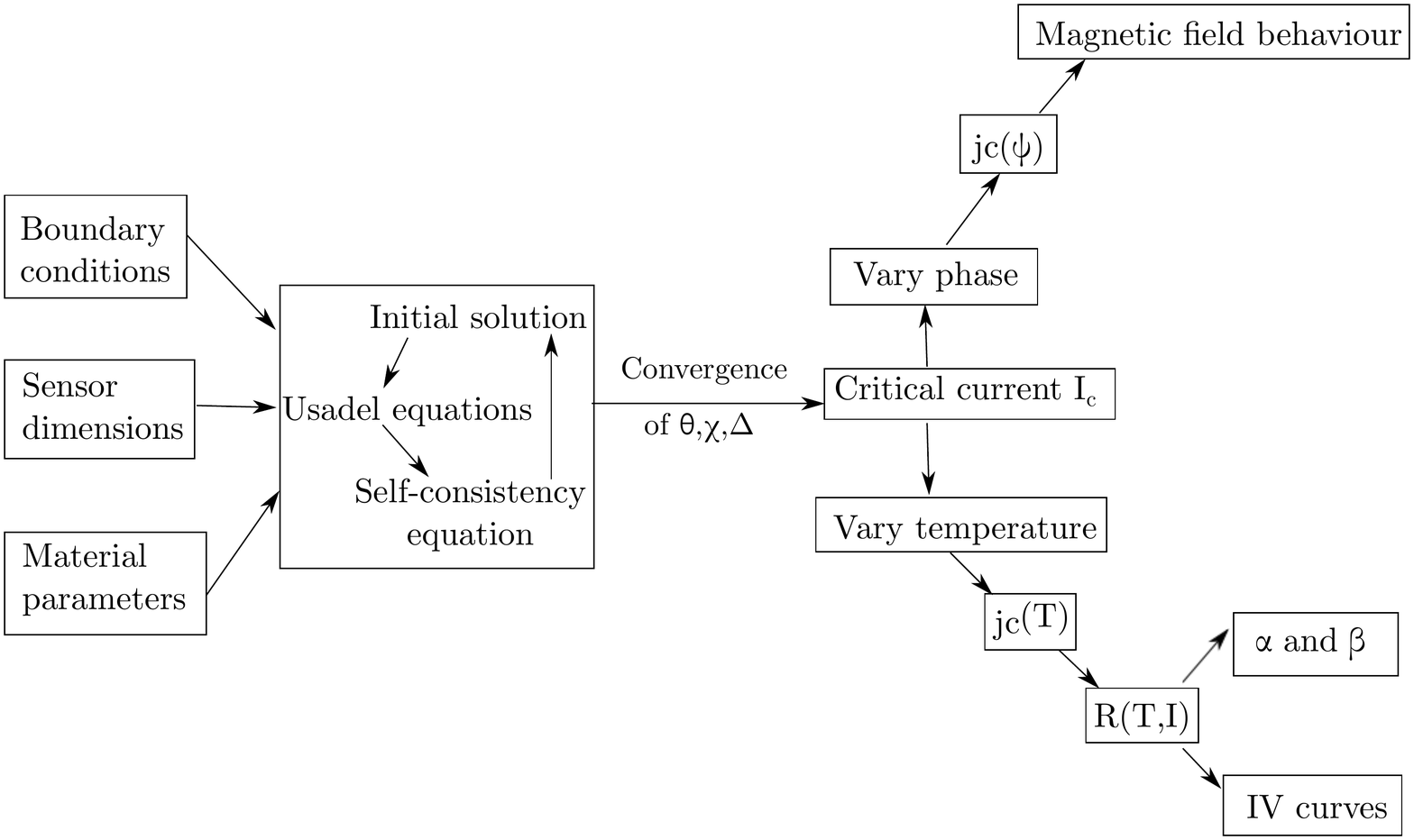}
\caption{\label{fig:modscheme} Diagram illustrating the process of finding the $I-V$ characteristic curves, small-signal electrothermal parameters and magnetic field behaviour for a TES starting from the sensor dimensions and the materials used, along with appropriate boundary conditions. These are used to find an initial solution, which is used to solve the Usadel equations for $\theta$, the pairing angle, and $\psi$, the superconducting phase. The superconducting order parameter $\Delta$ can then be found using the self-consistency equation and the initial solution is updated. This process is iterated until convergence of $\theta$, $\phi$ and $\Delta$ is achieved, and the critical current can then be calculated. Varying the superconducting phase gives information about the magnetic field behaviour of the TES; varying the temperature allows the calculation of $R(T,I)$, from which $I-V$ curves can be calculated.}
\end{figure}

Using a proximity effect model of a TES we can calculate parameters that can be compared with experimental measurements, and the procedure for this is shown in Figure \ref{fig:modscheme}. Our model \cite{Harwin2017} is based on the numerical solution by iteration of the Usadel equations \cite{Kozorezov2011, Kozorezov2011a} in conjunction with the self-consistency equation, subject to appropriate boundary conditions. \cite{Kuprianov1988} The other inputs to the model are the sensor dimensions and the material parameters. We then calculate the critical current density of the device for a particular value of superconducting phase difference $\psi$ and temperature $T$. Varying the superconducting phase allows us to investigate the magnetic field behaviour of the critical current; varying the temperature enables the calculation of $R(T,I)$ using the resistively shunted Josephson junction (RSJ) model. \cite{Kozorezov2011a,Coffey2008} From $R(T,I)$, the small signal electrothermal parameters $\alpha$ and $\beta$ and the $I - V$ characteristics can be calculated.

\begin{figure}
\centering
\includegraphics[width=0.5\textwidth]{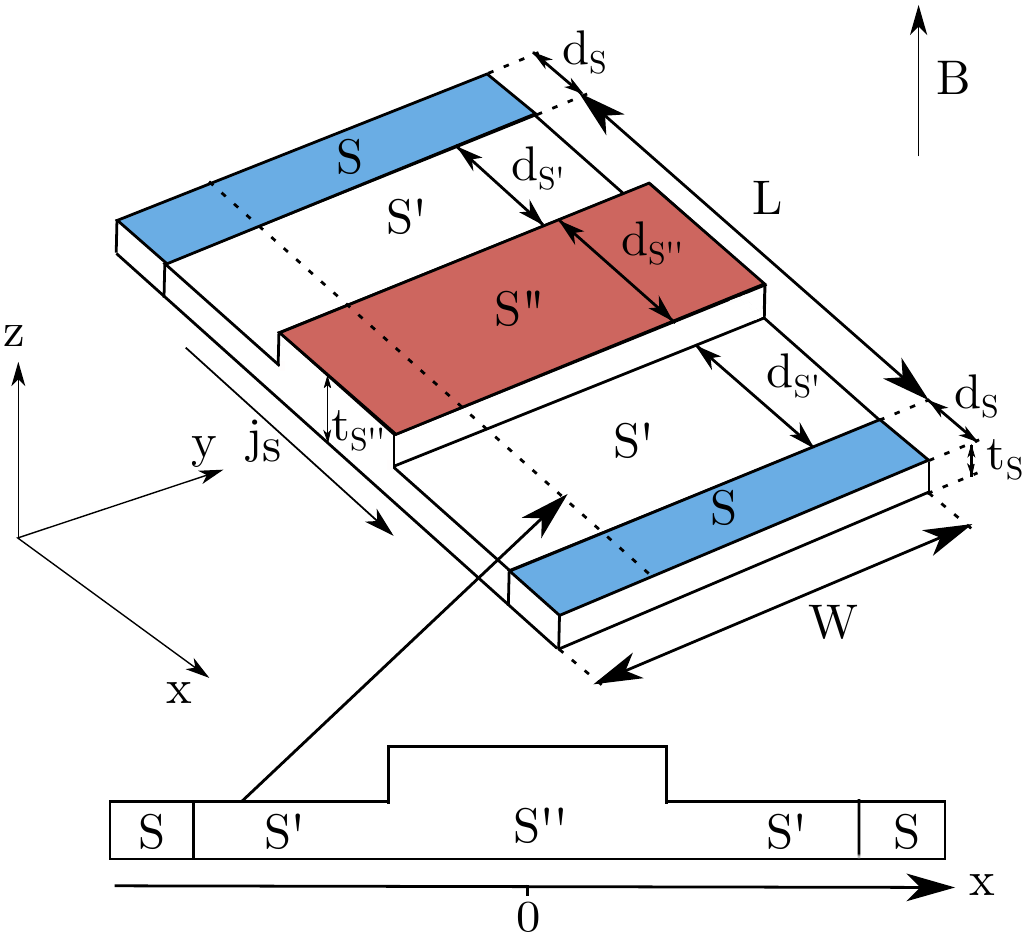}
\caption{\label{fig:TESdiagram} Geometry used to set up a proximity model of a TES, shown here for a TES with one normal metal bar. The axes indicate the coordinate system used to describe the sensor and the cut-through shows the reduction to the one-dimensional model. The TES is modelled as a series of one-dimensional slices, and the critical current densities in each slice are summed to calculate the overall critical current. We assume here that the supercurrent density $j_s$ is entirely in the $x$ direction. The (blue) regions labelled S are the superconducting electrodes; the unshaded regions labelled S' are the unpatterned areas of the bilayer and the (red) region labelled S'' is the normal metal bar. $L$ and $W$ are the length and width of the bilayer, respectively. $d_i$ is the width of region $i$ and $t_i$ its thickness. If a magnetic field is applied it is assumed to be perpendicular to the bilayer as indicated.}
\end{figure}

Figure \ref{fig:TESdiagram} shows the geometry used in this proximity model for a TES with a single normal metal bar. The $z$ axis is perpendicular to the device plane, the $x$ axis is along the length of the device and the $y$ axis is along the width. The regions labelled $\Sl$ are the superconducting electrodes; the regions labelled $\Sp$ are the unpatterned areas of the bilayer and the region labelled $\Spp$ is the normal metal bar. Each region i has width $d_i$ and thickness $t_i$. The bilayer has overall length $L$ and width $W$. We can vary the geometry used in the model to explore the effects of any additional normal metal structures such as bars

We model the TES in one dimension as shown by the cut-through in Figure \ref{fig:TESdiagram}, making the following assumptions:
\begin{enumerate}
\item{The electron diffusion in the $y$ direction is negligible, meaning that the drift current is parallel to the $x$ axis.}
\item{Any pair breaking by the current can be ignored.}
\item{In the $z$ direction, the superconducting film is thin enough ($t \ll \xi_{\mathrm{S'',S'}}$) that its properties are spatially invariant, even though the film is a proximity bilayer. We denote the composite bilayer simply as S', S'' as shown in Figure \ref{fig:TESdiagram}, and we include the effects of thickness changes.}
\item{In the y direction we assume that in the absence of a magnetic field the film is uniform, thereby reducing the problem to the $x$ direction.}
\item{The boundary conditions contain a parameter $\gamma_{\mathrm{B}}$, which is a measure of the boundary resistance at an interface between materials and which has not yet been measured experimentally. It is zero for a perfect interface and infinite for a completely insulating interface. In our modelling here we set $\gamma_{\mathrm{B(S,S')}} = 10$ at the S-S' boundary, giving a boundary resistance of 11.5 m$\Omega$. We take the interface between S' and S'' to have no boundary resistance, setting $\gamma_{\mathrm{B(S',S'')}} = 0$.}
\end{enumerate}

The equations to be solved are the one-dimensional Usadel equations, parameterised by the pairing angle $\theta$ and the superconducting phase $\chi$ \cite{Vasenko2008,Kozorezov2011,Golubov2004},
\begin{subequations}
\label{eq:finalus}
\begin{align}
\hbar D &\left(\frac{\Diff2 \theta}{\diff x^2} - \left(\frac{\diff \chi}{\diff x} \right)^2 \cos \theta \sin \theta \right)
= 2\hbar \omega \sin \theta - \cos \theta (\Delta \rme^{-\rmi \chi} + \Delta^* \rme^{\rmi \chi})  \label{eq:finalus1}, \\
\hbar D & \frac{\diff}{\diff x} \left(\sin^2 \theta \frac{\diff \chi}{\diff x} \right) = \rmi(\Delta \rme^{-\rmi\chi} - \Delta^* \rme^{\rmi\chi}) \sin \theta \label{eq:finalus2}.
\end{align}
\end{subequations}
Here $D$ is the diffusion coefficient of the material, $\Delta$ is the order parameter and $\omega$ represents the Matsubara frequencies, given by $\hbar \omega = \pi k_{\mathrm{B}} T (2n+1)$ for integer $n$. The Usadel equations are solved for $\theta$ and $\chi$ in conjunction with the self-consistency equation, which describes the variation of the order parameter
\begin{equation}
\label{eq:final1d}
\Delta \ln \left(\frac{T}{T_{\crit}} \right) + 2 \pi k_{\mathrm{B}} T \sum_{\omega > 0} \left( \frac{\Delta}{\hbar \omega} - \sin \theta \rme^{\rmi \chi} \right) = 0,
\end{equation}
with $T_{\crit}$ the transition temperature of the material. 

These three equations must be solved subject to suitable boundary conditions, as given by Kuprianov and Lukichev \cite{Kuprianov1988}, to ensure the conservation of the current at any material interfaces. We assume bulk superconductor behaviour at the external boundaries $x = \pm (L/2 + d_{\Sl})$; the outer edges of the superconducting electrodes. We solve the equations iteratively as shown in Figure \ref{fig:modscheme}, to obtain the values of $\theta(x), \chi(x)$ and $\Delta(x)$. From this the supercurrent density can be calculated,
\begin{equation}
\label{eq:finaljs}
j_{\mathrm{s}}(x,\psi) = -\frac{2\sigma \pi k_{\mathrm{B}} T}{e} \sum_{\omega >0} \sin^2 \theta \frac{\diff \chi}{\diff x},
\end{equation}
where $\psi = \chi(L/2+d_{\Sl})-\chi(-(L/2+d_{\Sl}))$ is the phase difference between the superconducting leads. $\psi$ depends on magnetic field according to $ \psi(y) = 2 \pi |\mathbf{B}|  (L+2 \lambda_{L}) y/ \Phi_0 + \psi_0$, where the magnetic field $\mathbf{B}$ has direction as shown in Figure \ref{fig:TESdiagram}. $\lambda_{L}$ is the penetration depth in the superconducting electrodes $\Sl$ and $\Phi_0 = h/2e$ is the quantum flux. $\psi_0$ is the value of $\psi$ that maximises the supercurrent in the absence of a magnetic field.

We define the critical supercurrent density as $j_{\crit} (x_m,\psi)$, where $x_m$ is the value of $x$ that minimises $(j_s(x,\psi))$. With zero applied magnetic field, $\psi = \psi_0$ does not vary with $y$ and the total critical supercurrent 
\begin{equation}
\label{eq:finalIs}
I_{\crit} (x_m, \psi_0) = W t(x_m)j_{\crit}(x_m, \psi_0)
\end{equation}
follows directly by integration of the supercurrent density with respect to $y$.

By varying the temperature $T$ at which we solve the Usadel equations, we can find the variation of the critical current with temperature, $I_{\crit}(T)$. From this, we calculate the TES resistance as a function of temperature $T$ and current $I$ using a RSJ model. The resistance is given by
\begin{equation}
\label{eq:resistance}
R(T,I) = R_{\nn} \left\{ 1 + \frac{1}{\kappa} \Im \left[ \frac{\mathcal{I}_{1+\rmi \zeta \kappa}(\zeta)}{\mathcal{I}_{\rmi \zeta \kappa}(\zeta)} \right] \right\},
\end{equation}
where $\zeta = \hbar I_{\crit}/2eT$ describes the effect of thermal fluctuations and $\mathcal{I}_{\mu}(\nu)$ are modified Bessel functions of the first kind of complex order $\mu$ and real variable $\nu$. The ratio of current to critical current is $\kappa = I/I_{\crit,T}$. The Bessel functions are calculated using a continued fraction method \cite{Gautschi2016}.

Once we know $R(T,I)$ we can calculate $I-V$ curves. The two equations that determine the $I-V$ characteristics of a TES are Ohm's law,
\begin{equation}
\label{eq:IVOhm}
V = IR(T,I),
\end{equation}
and the power balance equation, which equates the electrical power dissipated in the TES with the heat flow to the thermal sink,
\begin{equation}
\label{eq:IVtherm}
\frac{V^2}{R(T,I)} = K(T^n-T_b^n).
\end{equation}
Here $V$ is the voltage across the device, $T_b$ is the temperature of the thermal sink and $K$ and $n$ are parameters that characterise the temperature dependence of the thermal conductance, obtained from a fit to experimental data. As the analytical form of R(T,I) includes Bessel functions, we solve these equations using the following numerical procedure:

\begin{enumerate}
\item{Calculate $V-IR(T,I)$ and $V^2 - R(T,I)K(T^n-T_b^n)$, where $R(T,I)$ is a matrix calculated using Equation \eqref{eq:resistance}. This produces two matrices.}
\item{Find the zero crossings in each matrix, which define two lines in the $I-T$ plane. One equation is satisfied on each line.}
\item{Find the intercept of these zero crossing lines, which will be the point at which both Equations \eqref{eq:IVOhm} and \eqref{eq:IVtherm} are simultaneously satisfied.}
\item{Record the $I$ and $T$ values of this point and repeat for a series of voltages.}
\end{enumerate}

We repeat this procedure for each bath temperature to obtain a series of $I-V$ curves.

\begin{figure}
\centering
\includegraphics[width=0.5\textwidth]{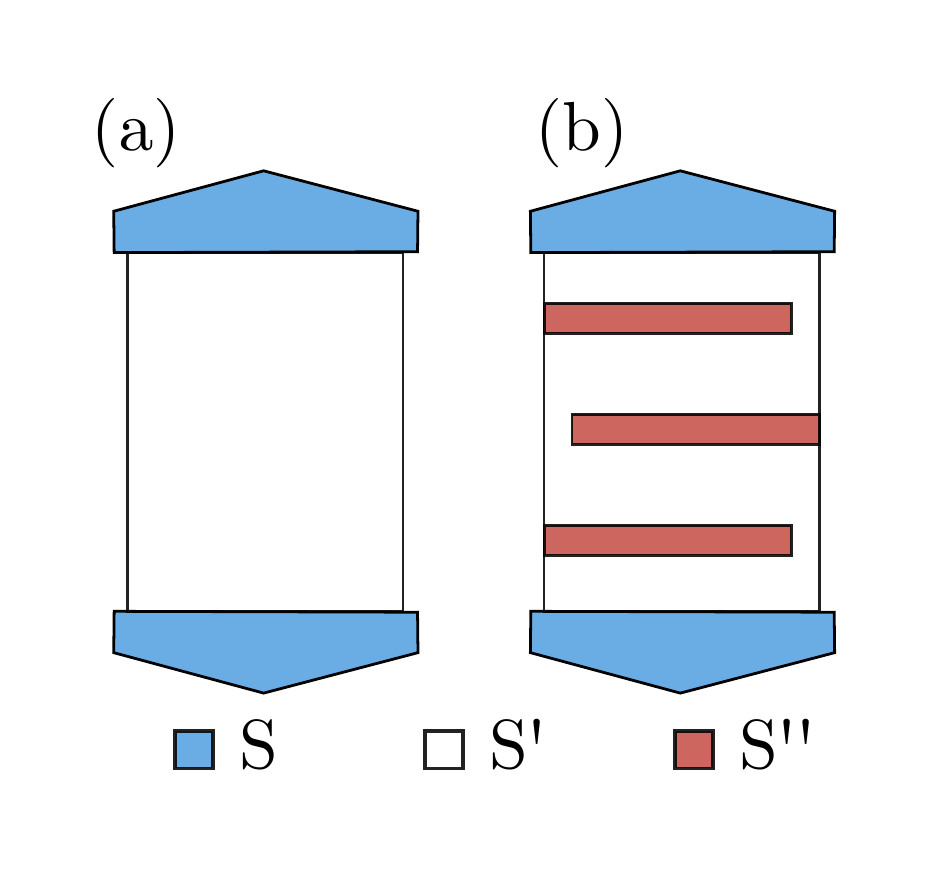}
\caption{\label{fig:devcomp} Device designs are shown here in top-down view. Both TESs have tapered electrodes ($\Sl$). Device 1 is of design (a), with no normal metal patterning. Device 2 has three partial normal metal bars ($\Spp$), as in design (b).}
\end{figure}

To see how effectively our model could describe the behaviour of actual devices, we applied this model to calculate $I-V$ curves for two TESs being developed as part of the Athena programme. We refer to these devices here as Device 1 and Device 2. Device 1 is a Ti/Au TES with no normal metal patterning, whilst Device 2 has three partial gold bars across the width of the device. Top-down views of the designs are shown in Figure \ref{fig:devcomp}. The key device parameters are given in Table \ref{tab:devdesc}. Both TESs are based on a Ti/Au bilayer (S') of thickness 19/70 nm, with $W$ = 100 $\mu$m and $L$ = 140 $\mu$m, and have a normal resistance of approximately 250 m$\Omega$. The additional gold bars (S'') have thickness 50 nm. This is the first time that Ti/Au devices with and without normal metal bars have been compared theoretically.

\begin{table}
\centering
\caption{\label{tab:devdesc} Key device parameters for the two devices under study. Both devices have a normal state resistance of 250 m$\Omega$. The unpatterned bilayer (S') consists of 19/70 nm of Ti/Au, has coherence length $\xi$ = 0.94 $\mu$m and dimensions $W$ = 100 $\mu$m, $L$ = 140 $\mu$m.}
\begin{tabular}{| l | c | c |}
\hline
 & Device 1 & Device 2 \\
\hline
Design type & (a) & (b) \\
Number of additional Au bars & 0 & 3 \\
Length of Au bars/$\mu$m & n/a & 90 \\
Thickness of Au bars/nm & n/a & 50 \\
Width of Au bars/$\mu$m & n/a & 5 \\
Transition temperature/mK & 102.4 & 101.7 \\
n, the exponent in the power balance equation & 3.63 & 3.62 \\
\hline
\end{tabular}
\end{table}

\section{RESULTS}

\begin{figure}
\centering
\begin{subfigure}{0.45\textwidth}
\centering
\includegraphics[width=\textwidth]{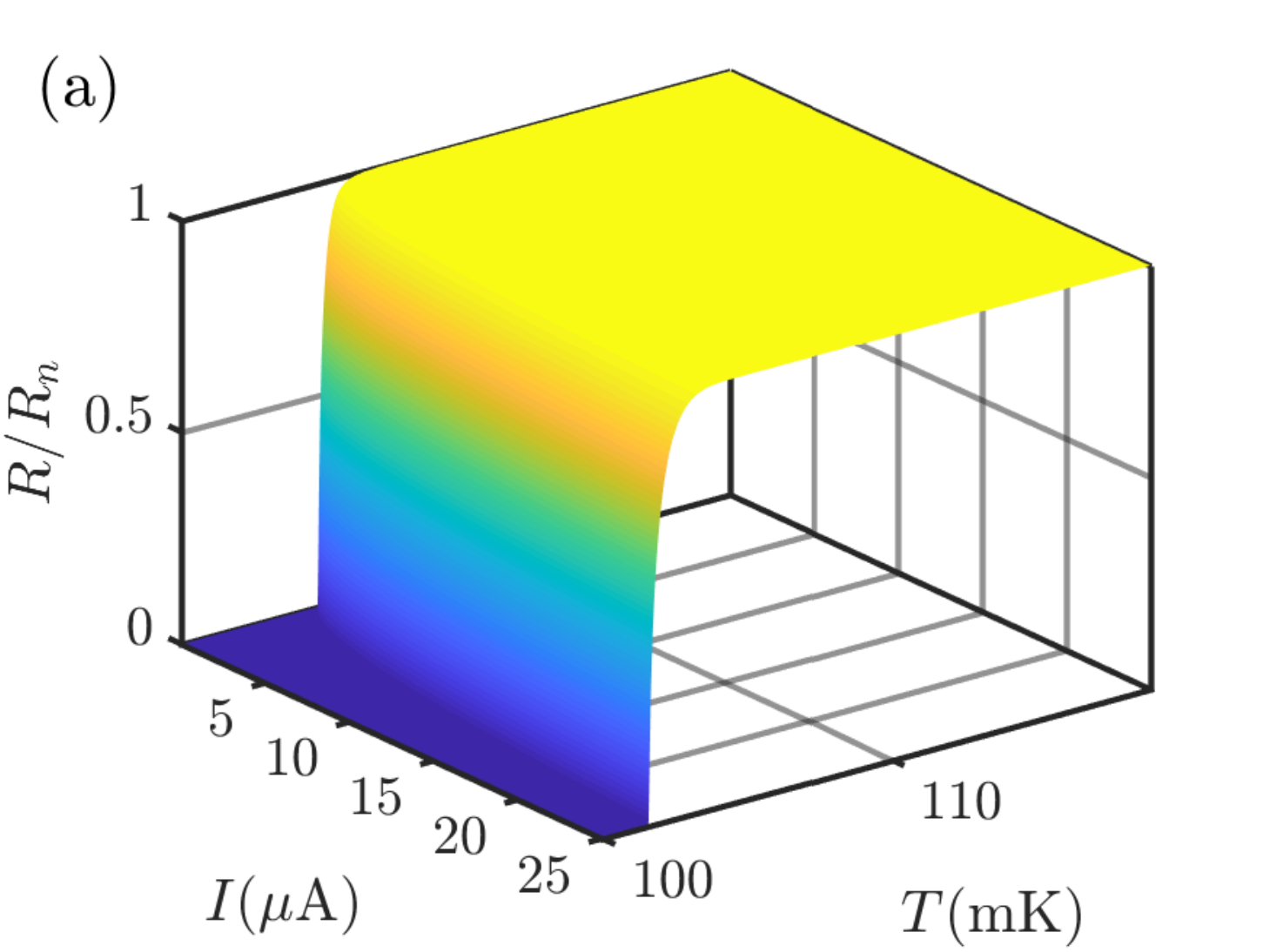}
\end{subfigure}
\hfill
\begin{subfigure}{0.45\textwidth}
\centering
\includegraphics[width=\textwidth]{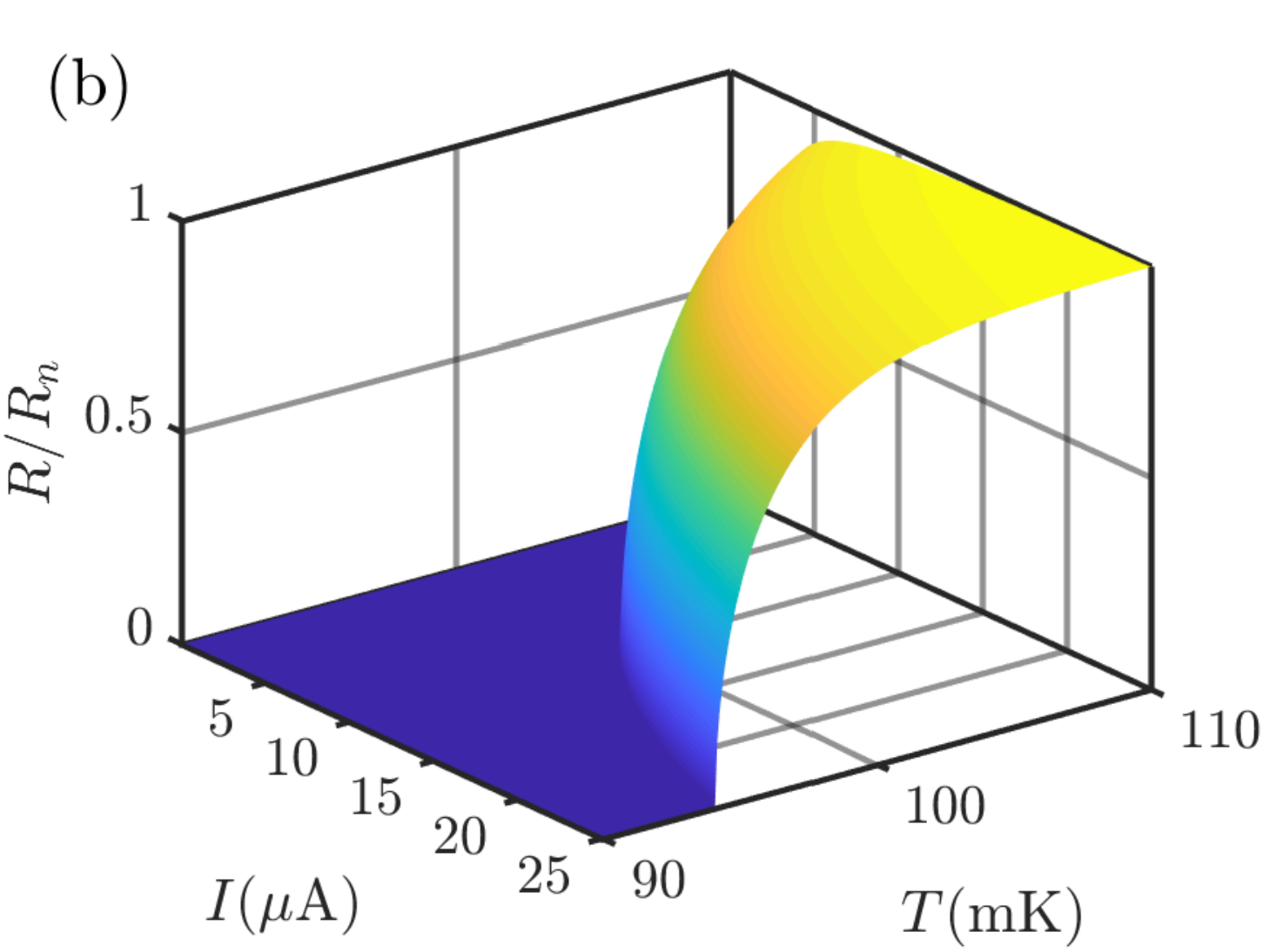}
\end{subfigure}
\caption{\label{fig:RTIs} $R(T,I)$ surfaces calculated using the proximity effect model for (a) Device 1 and (b) Device 2. Resistance is normalised to the normal state resistance of the device $R_{\mathrm{n}}$. Note that for both devices, the temperature range shown is $\Delta T$ = 20 mK.}
\end{figure} 

Figure \ref{fig:RTIs} shows the calculated $R(T,I)$ surfaces for Device 1 and Device 2, over a temperature range $\Delta T$ = 20 mK. It can be seen that Device 2, the sensor with normal metal bars, has a much broader transition than Device 1, the plain sensor, which indicates that the metal bars broaden the transition, reducing the value of  $\alpha$ of the device. Additionally, the presence of the metal bars lowers the initial onset of resistance from just above 100 mK to about 93-94 mK at a current of 25 $\mu$A. For lower currents, the transition temperature, defined as the temperature at which the device resistance is 50 \% of its normal state value, is approximately the same for both devices, which agrees with experimental observations.

\begin{figure}
\centering
\begin{subfigure}{0.45\textwidth}
\centering
\includegraphics[width=\textwidth]{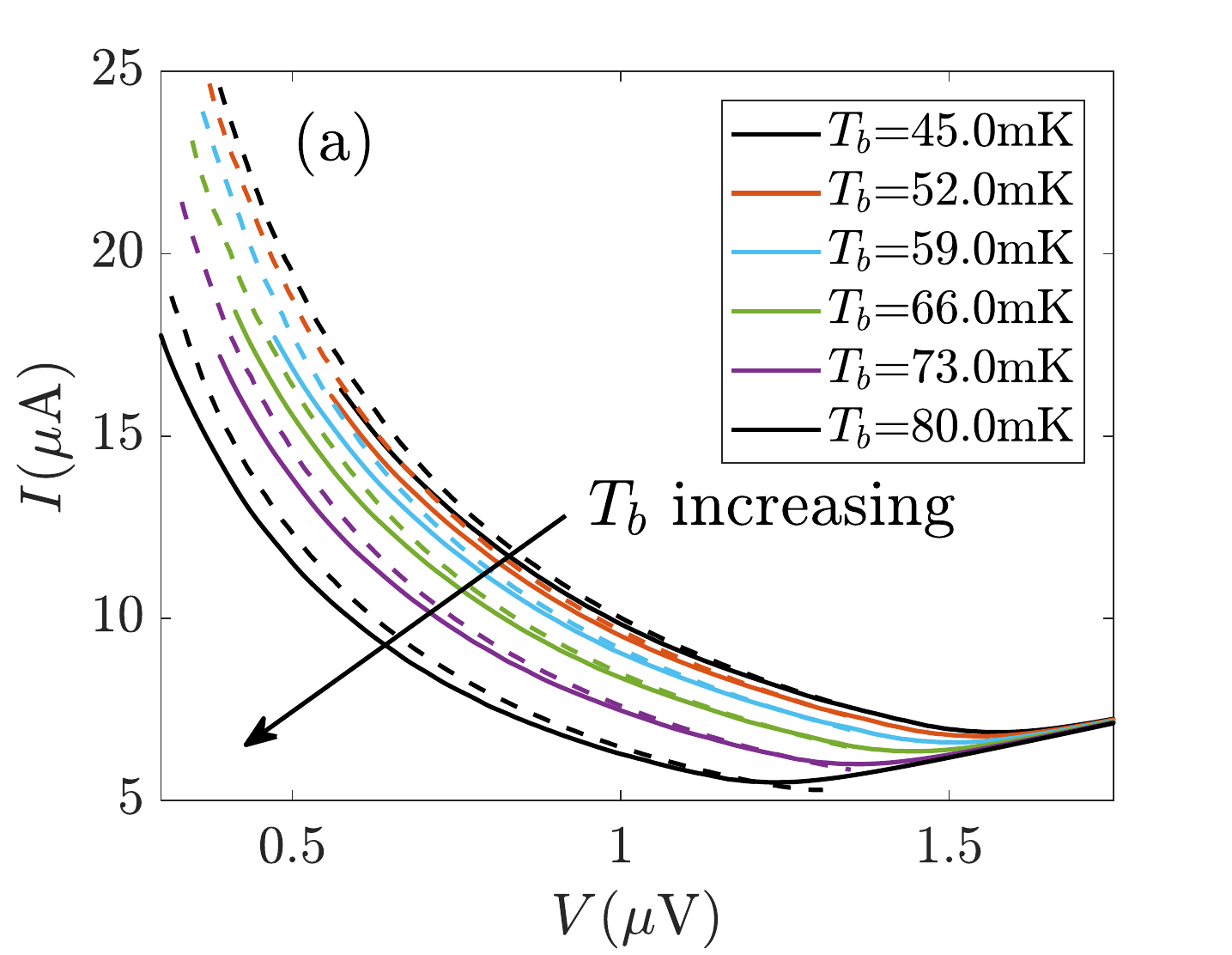}
\end{subfigure}
\hfill
\begin{subfigure}{0.45\textwidth}
\centering
\includegraphics[width=\textwidth]{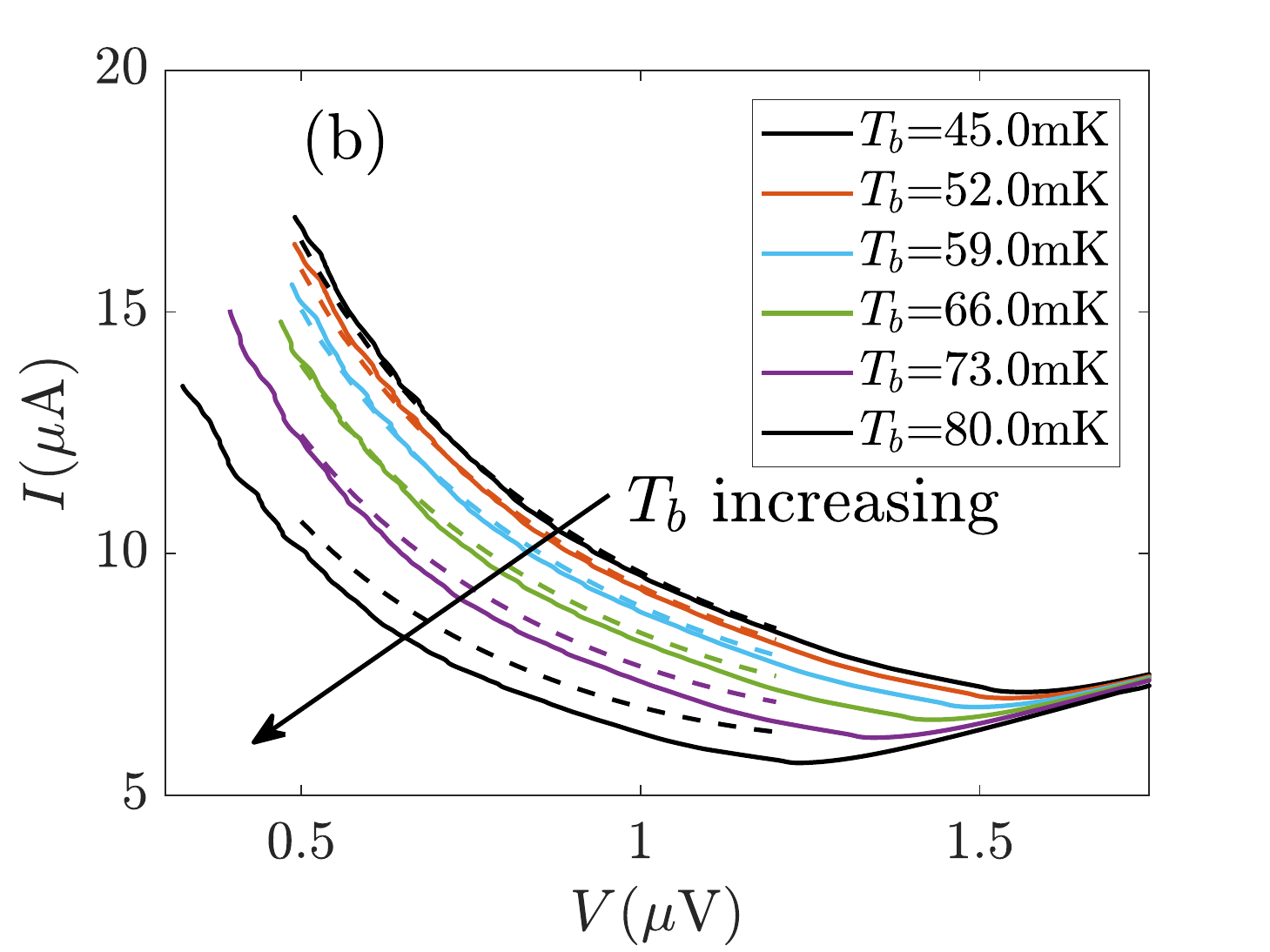}
\end{subfigure}
\caption{\label{fig:SRONIVs} $I-V$ curves as measured (solid lines) and calculated using the proximity effect model (dashed lines) over a range of bath temperatures $T_b$. The normal metal bars on Device 2 have an effective transition temperature $T_{\mathrm{c},S''} = 0.67$\:$  T_{\mathrm{c},S'}$}
\end{figure}

In Figure \ref{fig:SRONIVs}(a), it can be seen that the $I-V$ curves calculated using the proximity effect model agree extremely well with the $I-V$ curves measured experimentally for the plain TES, Device 1. The measured $I-V$ curves are shown as solid lines with the corresponding calculated $I-V$ curves shown as dashed lines, for a range of bath temperatures. The model correctly predicts both the shape of the $I-V$ curves and their variation with bath temperature. It should be emphasised that no attempt has been made to fit the model to the data. There is one free parameter in the modelling, $\gamma_{\mathrm{B}}$, which is a measure of the interface resistance between the niobium (S) electrodes and the Ti/Au (S') bilayer. Here we have taken $\gamma_{\mathrm{B}} = 10$ for all bath temperatures, showing that the same value of the boundary resistance gives good agreement between the model and the experimental data for a variety of different bath temperatures.

Figure \ref{fig:SRONIVs}(b) shows the measured (solid lines) and calculated (dashed lines) $I-V$ curves for the device with three normal metal bars, Device 2. The agreement is very good, although it should be noted here that the transition temperature of the S'' region is not always known from experiments. For example, in Device 2, there is a thin (3 nm) adhesive layer of titanium between the bar and the bilayer. In this device we predict the transition temperature of the regions with normal metal bars $T_{\mathrm{c,S''}}$ to be equal to $0.67 \: T_{\mathrm{c},S'}$, based on the $I-V$ curves at $T_b$ = 45 mK. This transition temperature is taken to be the same for all of the different bath temperatures. We have again taken $\gamma_{\mathrm{B(S,S')}}$ = 10 for the interface between the niobium electrodes and the Ti/Au bilayer, and we assume that the resistance of the interface between the S'' region and the S' region is negligible so the interface parameter $\gamma_{\mathrm{B(S',S'')}}$ = 0. For both devices, the fact that the one-dimensional model does well at predicting the $I-V$ curves implies that any two-dimensional effects are not significant. This means that the current in Device 2 is mostly flowing perpendicular to the electrodes and not taking a meandering path to avoid passing through the $\Spp$ regions, as has been previously suggested. \cite{Swetz2012a}

\begin{figure}
\centering
\begin{subfigure}{0.45\textwidth}
\centering
\includegraphics[width=\textwidth]{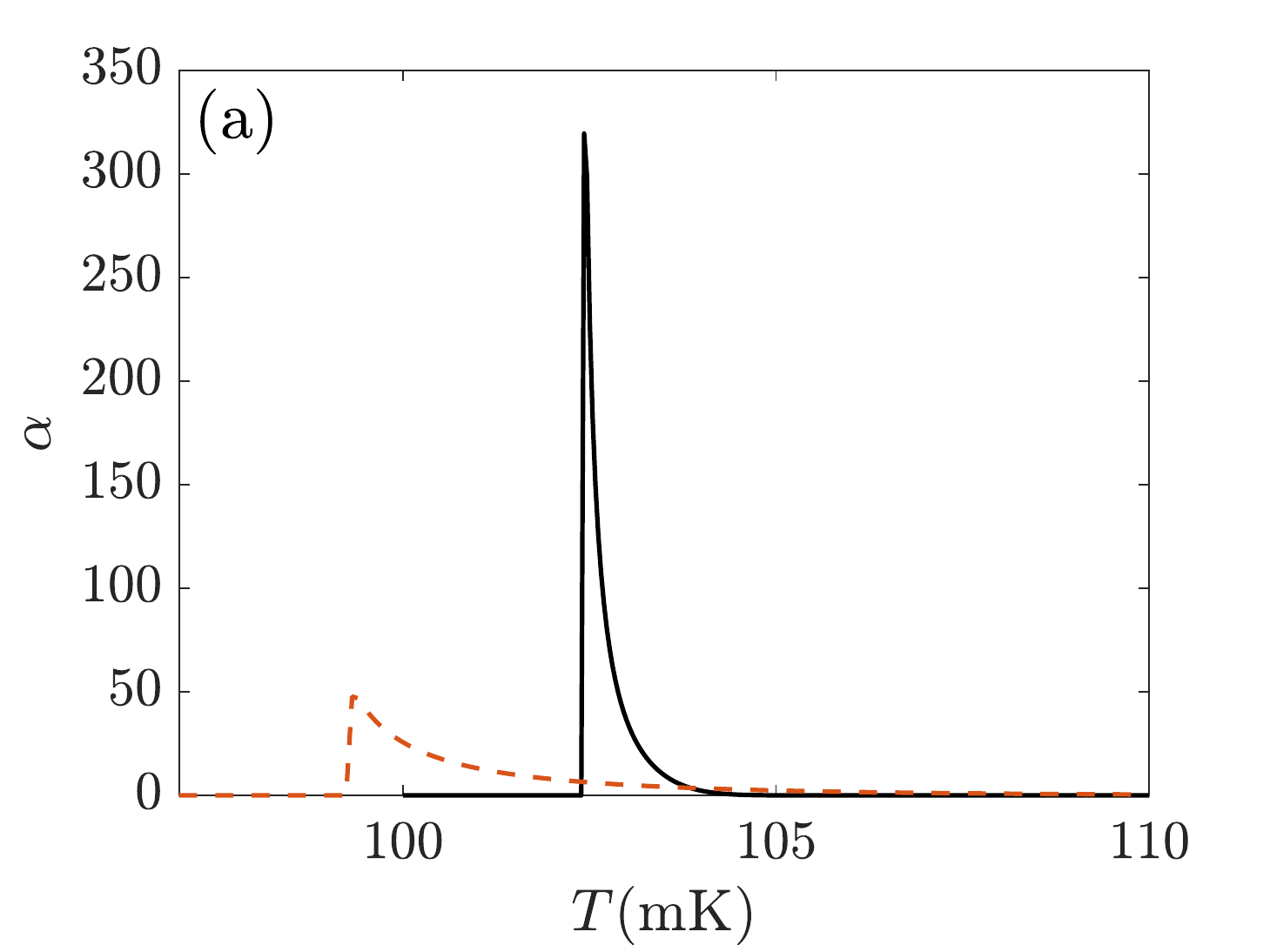}
\end{subfigure}
\hfill
\begin{subfigure}{0.45\textwidth}
\centering
\includegraphics[width=\textwidth]{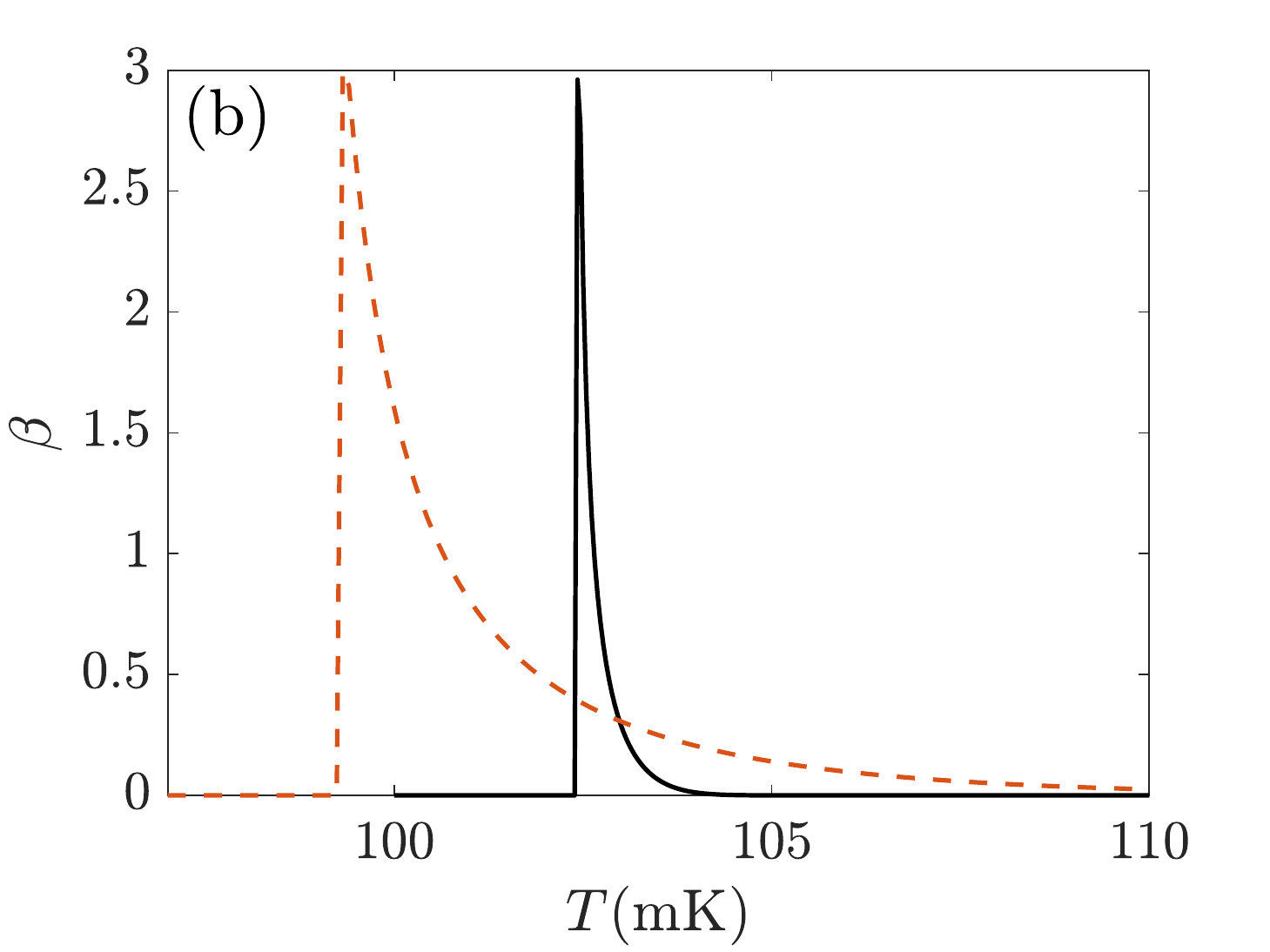}
\end{subfigure}
\caption{\label{fig:alphabet} Electrothermal parameters (a) $\alpha(T)$ and (b) $\beta(T)$ for Device 1 (solid lines) and Device 2 (dashed lines), calculated at a current of 12.5 $\mu$A.}
\end{figure} 

We have calculated the small signal electrothermal parameters $\alpha(T)$ and $\beta(T)$ for Device 1 and Device 2, as shown in Figure \ref{fig:alphabet}. The presence of the normal metal bars significantly reduces the maximum value of $\alpha$ by a factor of about 10, due to the broadening of the transition with temperature. The maximum value of $\beta$ is not significantly altered by the addition of normal metal bars, although the temperature broadening seen for Device 2, the device with normal metal bars, produces a slower decay of $\beta$.

\begin{figure}
\centering
\begin{subfigure}{0.45\textwidth}
\centering
\includegraphics[width=\textwidth]{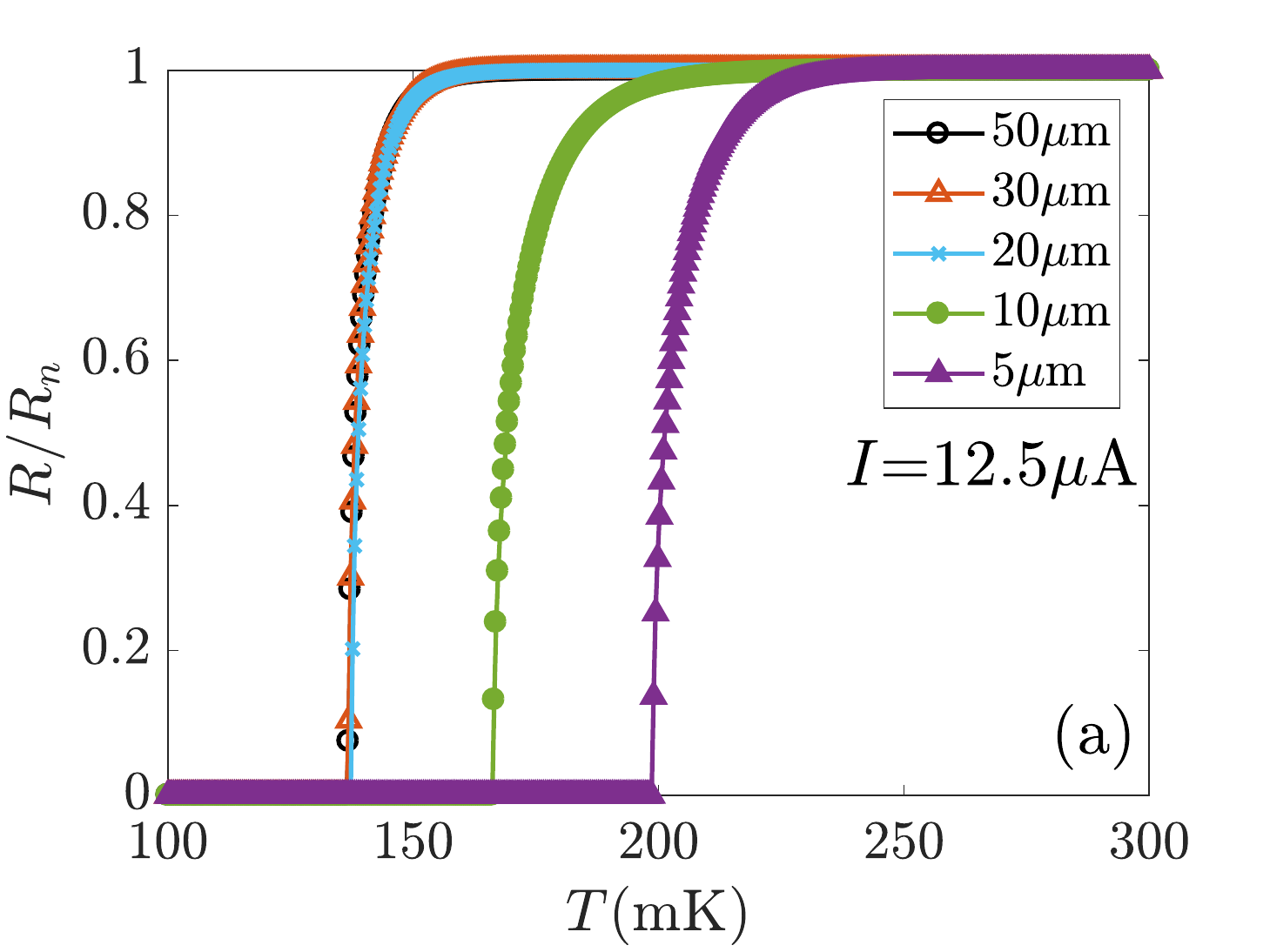}
\end{subfigure}
\hfill
\begin{subfigure}{0.45\textwidth}
\centering
\includegraphics[width=\textwidth]{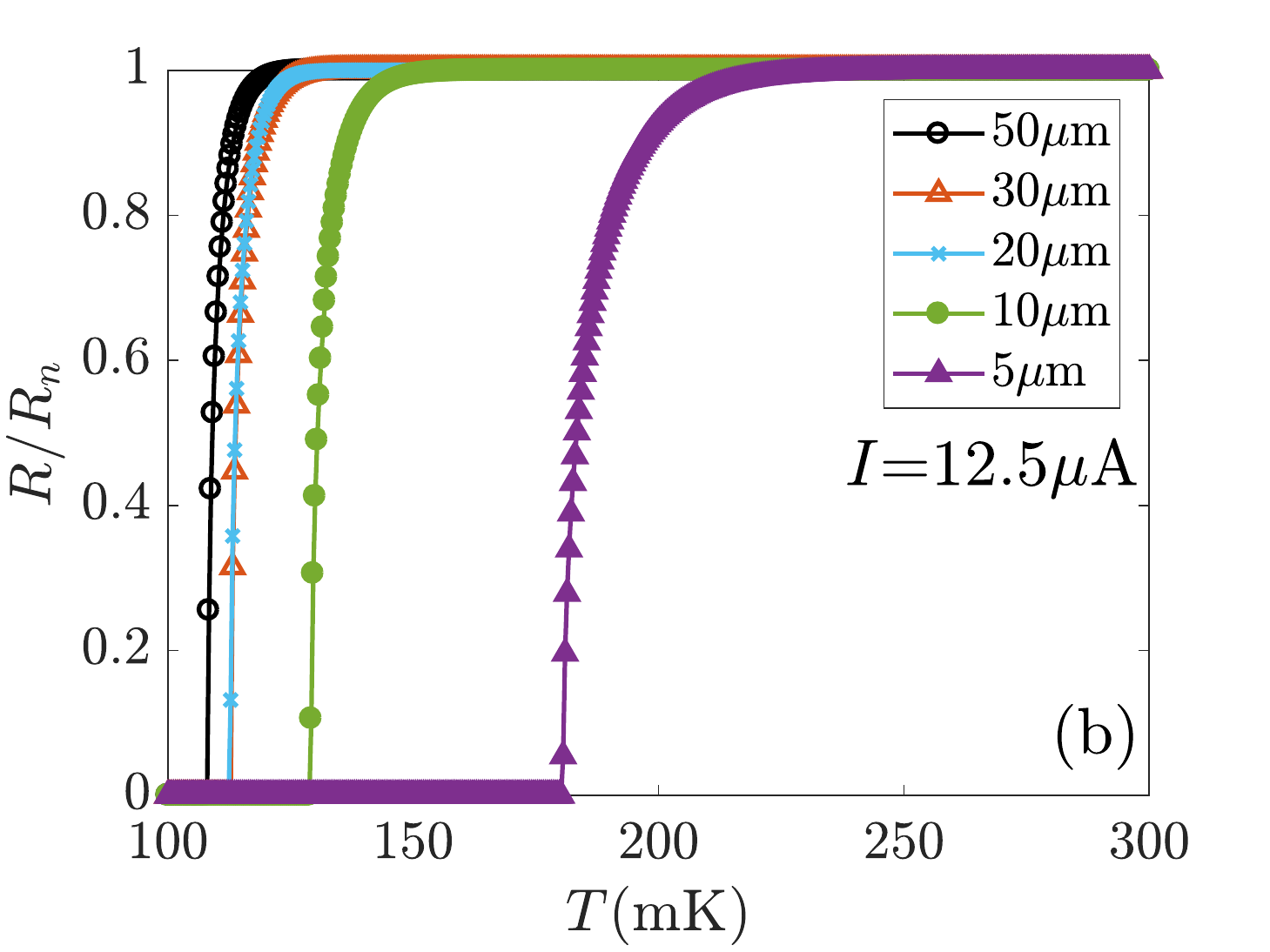}
\end{subfigure}
\caption{\label{fig:RTIsize} $R(T)$ curves calculated at 12.5 $\mu$A for Ti/Au TESs based on square bilayers with different side lengths. (a) is for plain TESs; (b) is for devices with three additional gold bars deposited on the bilayer. The bars are assumed to be long as the device is wide and their widths scale in proportion to the side length of the device.}
\end{figure} 

\begin{figure}
\centering
\begin{subfigure}{0.45\textwidth}
\centering
\includegraphics[width=\textwidth]{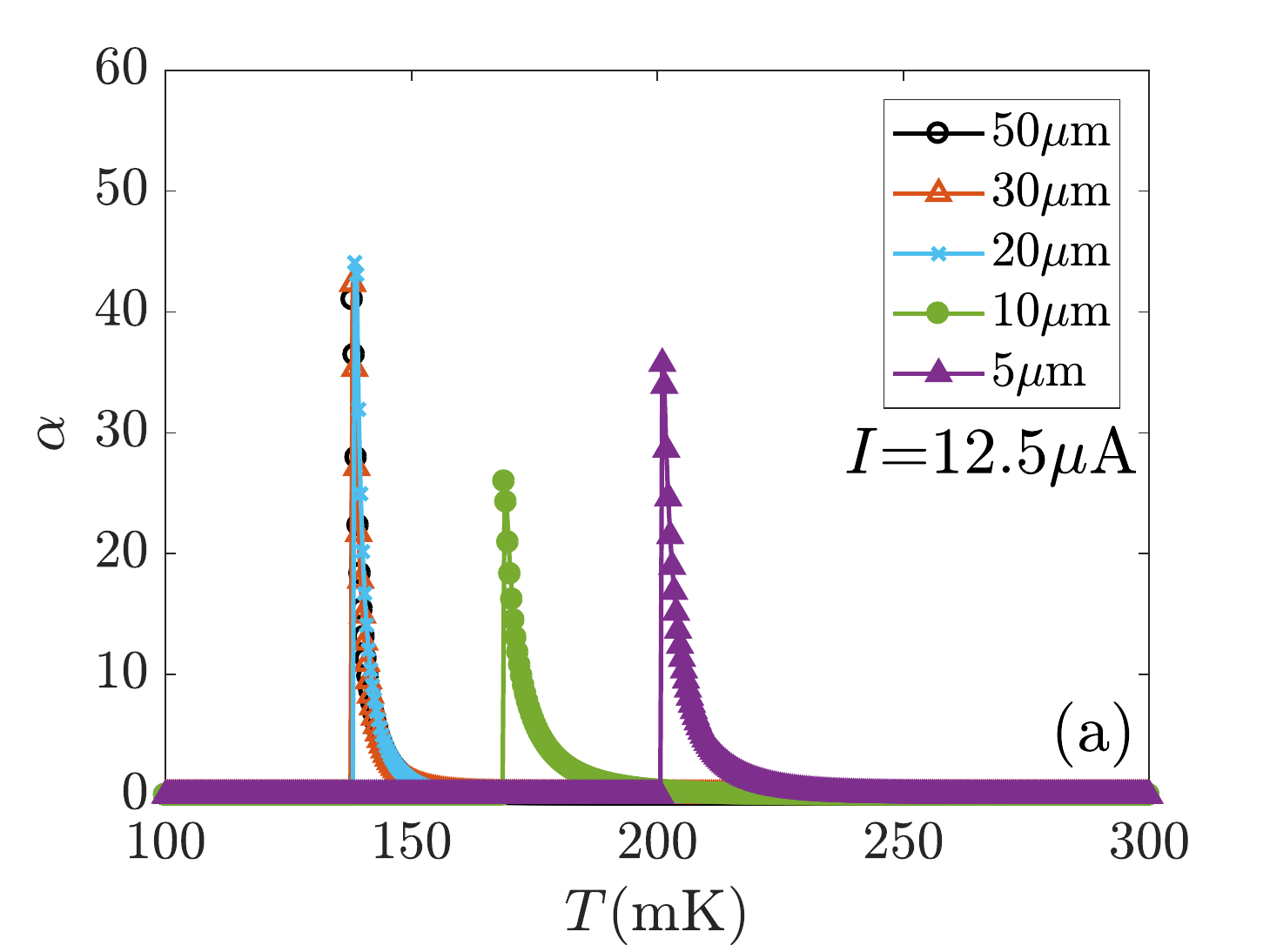}
\end{subfigure}
\hfill
\begin{subfigure}{0.45\textwidth}
\centering
\includegraphics[width=\textwidth]{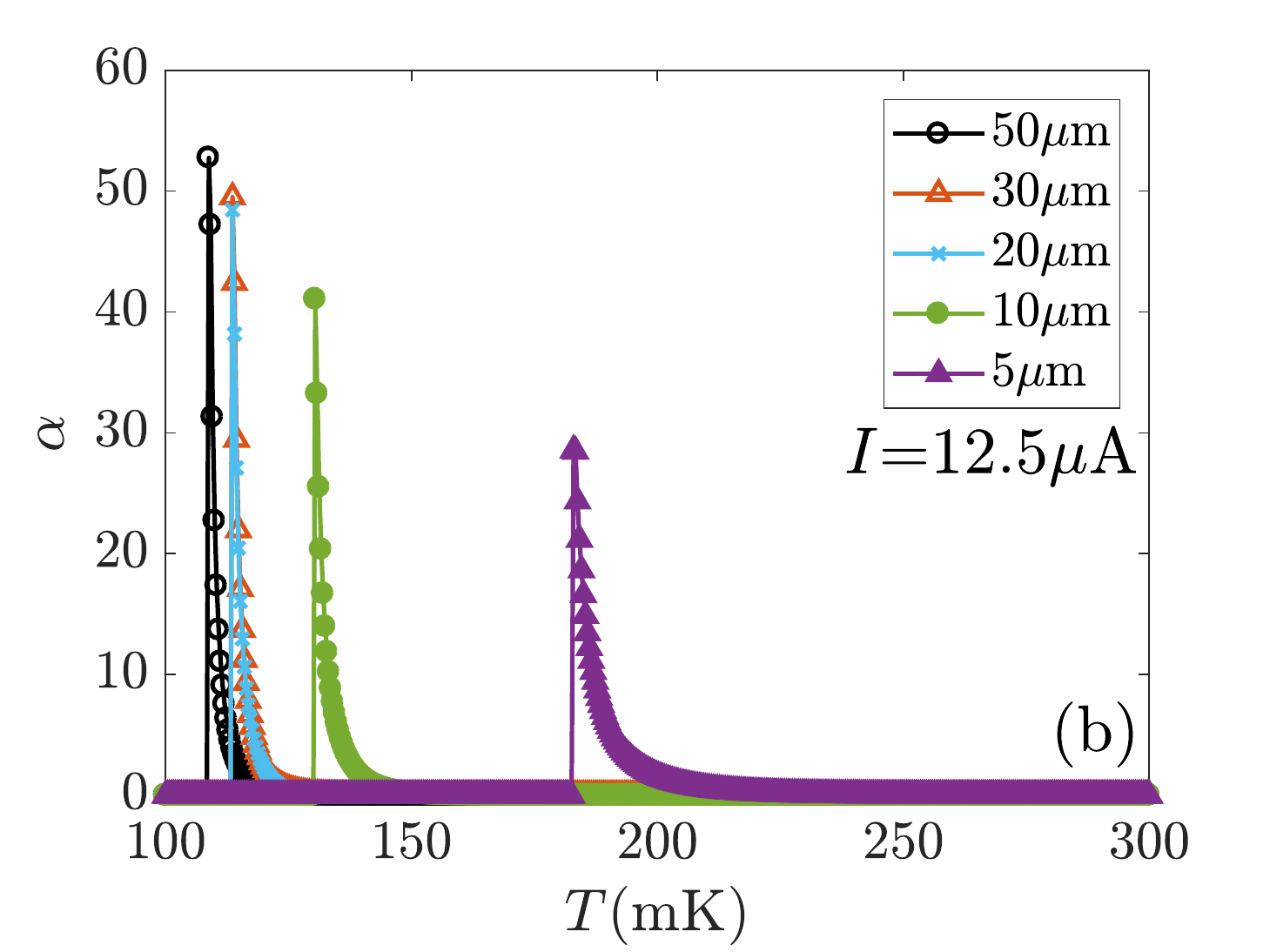}
\end{subfigure}
\caption{\label{fig:alphasize} $\alpha(T)$ curves calculated at 12.5 $\mu$A for Ti/Au TESs based on square bilayers with different side lengths. (a) is for plain TESs; (b) is for devices with three additional gold bars deposited on the bilayer. The bars are assumed to be as long as the device is wide and their widths scale in proportion to the side length of the device.}
\end{figure} 

We now use this model to investigate the effects of bilayer size on device performance. In Figure \ref{fig:RTIsize}, calculated $R(T)$ curves are shown for TESs based on square bilayers, with side lengths ranging from 50 $\mu$m to 5 $\mu$m. (a) is for bare devices; (b) is for devices with three normal metal bars. These bars are equally spaced and have widths proportional to the size of the device. There is not much difference in transition width or temperature between the 50, 30 and 20 $\mu$m devices. There is a significant increase in transition temperature from the 20 $\mu$m devices to the 10 $\mu$m devices, and a further increase in $T_{\crit}$ between the 10 $\mu$m devices and the 5 $\mu$m devices. As can be seen in Figure \ref{fig:alphasize}, the value of $\alpha$ does not significantly change with the addition of normal metal bars for these smaller devices. With the exception of the 5/10 $\mu$m plain devices, the maximum value of alpha reduces slightly as the device size increases. 

\begin{figure}
\centering
\includegraphics[width=0.5\textwidth]{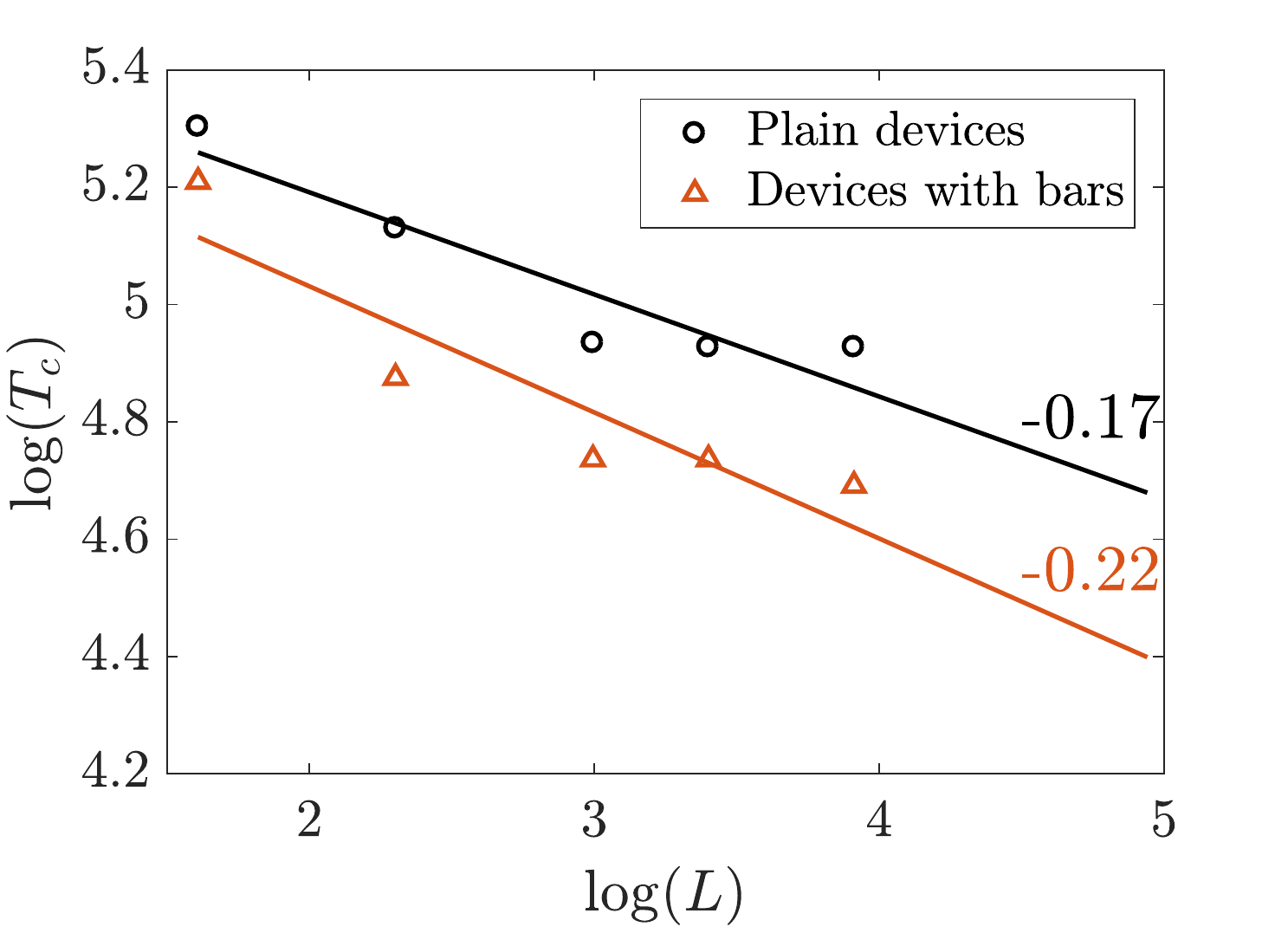}
\caption{\label{fig:TcL} $T_{\crit}$ as a function of device side length $L$. For a relationship of the form $T_{\crit} \propto L^{m}$, we find $m$ = -0.17 for plain devices (black plot with circles) and $m$ = -0.22 for devices with normal metal bars (red plot with triangles). The solid lines show the fits to the data which were used to find the values of $m$. The current is 12.5 $\mu$A.}
\end{figure}

We investigate the relationship between $T_{\crit}$ and $L$ in Figure \ref{fig:TcL} for devices both with and without normal metal bars. For a relationship of the form $T_{\crit} \propto L^{m}$, we find $m$ = -0.17 for plain devices and $m$ = -0.22 for devices with normal metal bars. These values of $m$ are different to those for measured devices \cite{Sadleir2010} and previous calculated relationships, \cite{Harwin2017} which may be due to the difference in material parameters. Previous relationships have been measured and calculated for Mo/Au devices whilst these relations are calculated for Ti/Au devices. This variation of transition temperature with device size could be exploited as another way of changing the transition temperature of a device, instead of altering the thicknesses of the bilayer components.

\begin{table}
\centering
\caption{\label{tab:devc} Estimates of heat capacity for devices based on square bilayers with different side lengths. All devices consist of a 17/70 nm Ti/Au bilayer, and any additional Au bars are 50 nm thick. We have assumed the devices do not have any absorbers attached.}
\begin{tabular}{| c | c | c |}
\hline
 Side length/ $\mu$m & C for device without bars/ fJ K$^{-1}$ & C for device with bars/ fJ K$^{-1}$  \\
\hline
50 & 3.01 & 3.86 \\
30 & 1.08 & 1.39 \\
20 & 0.48 & 0.62 \\
10 & 0.12 & 0.15 \\
5 & 0.03 & 0.04 \\
\hline
\end{tabular}
\end{table}

Table \ref{tab:devc} displays estimates of the heat capacities of square TES bilayers with different side lengths. Comparing the 50 $\mu$m device with the 5 $\mu$m device, the 50 $\mu$m device has a value of $\alpha$ about 1.5-2 times larger than the 5 $\mu$m device, and a heat capacity approximately 100 times larger. The smaller devices therefore contribute less to the overall heat capacity of the system. For TESs without absorbers, smaller devices will be faster, as the large reduction in heat capacity compared to the reduction in $\alpha$ shown in Figure \ref{fig:alphasize} leads to a decrease in $\tau_{\mathrm{eff}} \propto C/(G(1+\alpha/n))$. $G$ will not vary significantly for TESs with good thermal isolation.

We can draw conclusions about the optimal device geometry from the variation of transition temperature, $\alpha$ and heat capacity with device size. We should consider that for x-ray devices, the heat capacity $C$ will be dominated by the heat capacity of the additional absorber, and so will not vary significantly with device size. Therefore, for x-ray devices, the slightly higher $\alpha$ and lower transition temperature of larger devices makes larger devices more desirable in terms of speed and energy resolution. Although for these larger devices, metal bars reduce the value of $\alpha$ by up to an order of magnitude, they have also been shown to reduce unexplained noise. \cite{Ullom2004} This phenomenon is not described by our modelling, and so it should be considered whether or not the higher theoretical energy resolution for plain devices is counteracted by the additional unexplained noise. 

For devices without additional absorbers, smaller devices are more desirable, as they will have a faster response time and greater energy resolution. 10-20 $\mu$m square devices represent the best balance between a small heat capacity and a low transition temperature, and also provide good spatial resolution. These smaller devices do not show a significant reduction in $\alpha$ with the addition of normal metal bars, and so should be constructed with normal metal bars as this will provide a reduction in unexplained noise without a significant reduction of the theoretical energy resolution or response time.

\section{CONCLUSIONS}
\label{sec:conc}
We have demonstrated a numerical model of a TES based on the Usadel equations that successfully predicts $I-V$ curves for devices based on Ti/Au bilayers. We then used this model to investigate the effects of device geometry on performance for a series of smaller devices based on square Ti/Au bilayers with the same layup. However, we should note that this modelling does not currently account for the noise generated in the bilayer and at the moment, its dependence on device size is not entirely clear.

The choice of device size is important in a large number of different TES applications, from far infra-red sensors to optical detectors, although there are many other demands that are specific to the application of the TES. In general terms, the lateral proximity effect increases the device transition temperature for smaller devices. For devices with large absorbers such as x-ray TESs, larger devices with no additional normal metal bars provide better energy resolution and response time, but the bars can produce a reduction in unexplained noise. For devices without absorbers such as optical TESs, reducing the size of the device improves the response time whilst giving greater spatial resolution. For these smaller devices, there is no disadvantage to adding normal metal bars, as the bars reduce unexplained noise but do not significantly alter the value of $\alpha$. 
\acknowledgements
This work was partly supported by ESA CTP contract with No. 4000114932/15/NL/BW and EU H2020 AHEAD program.

\bibliography{/Users/astarael/Dropbox/PhD/Write-ups/library.bib} 
\bibliographystyle{spiebib} 

\end{document}